\documentclass[12pt]{article}
\usepackage{latexsym}
\usepackage{amssymb}
\usepackage{amsmath}
\usepackage{graphicx}
\newcommand{\pme}{^{\prime}}
\newcommand{\gam}{\gamma(r)_{\pm}}
\newcommand{\alp}{\alpha(r)_{\pm}}

\newcommand{\dgam}{\gamma(r)_{\pm,1}}
\newcommand{\smp}{\shortparallel}
\newcommand{\alphab}{\left(1-\frac{b_{\pm}(r)}{r}\right)}

\begin{document}

\title{On a General Class of Wormhole Geometries}
\author{{\small A. DeBenedictis
\footnote{e-mail: debened@death.phys.sfu.ca}}\\ \it{\small Department of
Physics} \\ \it{\small Simon Fraser University, Burnaby, British Columbia,
Canada V5A 1S6} \and {\small A. Das \footnote{e-mail: das@sfu.ca}} \\
\it{\small
Department of Mathematics and Statistics} \\ \it{\small Simon Fraser
University, Burnaby, British Columbia, Canada
V5A 1S6}}
\date{October 30, 2000}
\maketitle

\begin{abstract} A general class of solutions is obtained which describe a
spherically symmetric wormhole system. The presence of arbitrary functions
allows one to describe infinitely many wormhole systems of this type. The
source of the stress-energy supporting the structure consists of an
anisotropic brown dwarf ``star'' which smoothly joins the vacuum and may
possess an arbitrary cosmological constant. It is demonstrated how this
set
of solutions allows for a non-zero energy density and therefore
allows positive
stellar mass as well as how violations of energy conditions may be
minimized. Unlike examples considered thus far, emphasis here is placed on
construction by manipulating the matter field as opposed to the metric.
This scheme is generally more physical than the purely geometric method.
Finally, explicit examples are constructed including an example which
demonstrates how multiple closed universes may be connected by such
wormholes. The number of connected universes may be finite or infinite.
\end{abstract}

\vspace{3mm}

\section{Introduction}
\qquad The study of spacetimes with nontrivial topology is one with an
interesting history. One of the earliest, and most famous, wormhole
type geometries considered is that of the Einstein-Rosen bridge
\cite{ref:einstrose}. This ``bridge'' connecting two sheets was to
represent an elementary particle. That is, particles both charged and
uncharged were to be modeled as wormholes. From the work of Einstein
and Rosen one also can find a hint to what eventually was to become a
central problem in wormhole physics. Energy conditions must generally
be violated to support static wormhole geometries.

\qquad The wormhole resurfaced later in the study of quantum gravity
in the context of the ``spacetime foam'' \cite{ref:wheeler}
\cite{ref:MTW} in which wormhole type structures permeate throughout
spacetime at scales near the Planck length. More recently, there is
the meticulous study performed by Morris and Thorne
\cite{ref:morthorne} considering a static, spherically symmetric
configuration connecting two universes. Since the work by Morris and
Thorne, there has been a considerable amount of work on the subject of
wormhole physics. For an excellent review the reader is referred to
the thorough book by Visser \cite{ref:visbook} as well as the
exhaustive list of references therein. Most interesting is the case
of the traversable wormhole which, at the very
least, requires tidal forces to be small along time-like world lines
as well as the absence of event horizons. These geometries are of interest not only for
pedagogical reasons but they serve to shed light on some fundamental issues involving
chronology protection \cite{ref:hawkchron} and the topology of the universe. Also, many
wormhole solutions are in violation of the weak energy condition and there has been much
activity in the literature to attempt to minimize such violations (see, for
example, \cite{ref:vis1}, \cite{ref:delgaty},
\cite{ref:vis2}, \cite{ref:anchordoqui} and references therein.) There is also
the possibility that wormholes have much relevance to studies of black holes. For example, it
may be possible that small wormholes may induce large shifts in a black hole's event horizon
or that semiclassical effects near a singularity may cause a wormhole to form avoiding a
singularity altogether \cite{ref:froandnov}, \cite{ref:visrice},
\cite{ref:vishafia}.

\qquad For the above reasons, and many others which are hinted at in the
literature, it is instructive to study as general a class as possible of
such solutions. The aim here is to construct a mathematical model which
encompasses the bulk of static, spherically symmetric, wormhole solutions.
This is done by studying the profile curve of a generic wormhole structure
and postulating a mathematical expression describing this curve. From this
ansatz all relevant properties may be calculated. Here we are concerned
with making the system as physically relevant, yet as general, as
possible. This involves, as much as possible, working with the stress
energy tensor when studying the system, as opposed to fully prescribing
the geometry as is usually done. We demand that the physical requirements
be as reasonable as possible given the situation and study the system from
this point of view. Also, the system is not supported by a matter field
which permeates all space but possesses a definite matter boundary. At
this boundary the solution is smoothly patched to the vacuum with
arbitrary cosmological constant. The presence of a cosmological constant
is permitted for several reasons. First, astronomical studies of distant
supernovae favour the presence of a non-zero cosmological constant
\cite{ref:reiss} \cite{ref:perl}. Second, the cosmological constant
necessarily violates the strong energy condition and therefore may serve
to minimize energy condition violations of the matter field. We make
comments where appropriate on how the introduction of the cosmological
term affects the behaviour of the matter field. In this paper we find that
solutions may be obtained in which the matter field does not violate
energy conditions at the throat and with minimal energy condition
violation elsewhere. It is possible that a patch to an intermediate layer
of matter may eliminate violations altogether.

\qquad The paper is organized as follows: Section 2 introduces the
Einstein field equations for a general static spherically symmetric
system, as well as the differential identities which must be satisfied.
This
system is then specialized to describe a general wormhole system. Section
3 illustrates the methods used to solve the field equations. The matter
field is also introduced and restrictions on its behaviour are analyzed.
The singularity and event horizon conditions are studied here in terms of
the matter field. The smooth patching of this solution to the exterior
Schwarzschild-(anti) de Sitter solution \cite{ref:kottler} is also covered
here along with junction conditions. Specific examples are constructed and
energy conditions are studied in section 4. Finally, some concluding
remarks are made in section 5.

\section{Geometry and Topology}
\qquad The geometry studied is displayed in figure 1. For the moment we
restrict our analysis to the case where there exist two regions, an
upper and lower sheet, which are connected by a throat of radius
$r=r_{0}$. Neither region need be asymptotically flat and each region
may represent a universe which is connected by the wormhole throat. It
would not be difficult to modify the calculations here to represent a
throat connecting two parts of the same universe. This, for example,
may easily be accomplished via topological identification at spatial
infinity in the case of an open universe. The closed universe case will be
considered as a specific example later.

\qquad In the case studied here both the upper and lower universes
possess a matter-vacuum boundary at $r=d_{\pm}> r_{0}$ where the $+$
applies to the ``upper universe'' and the $-$ denotes quantities in the
``lower universe''. This boundary represents the junction between the
brown dwarf and the vacuum. We use the term brown dwarf to describe
the matter system for several reasons. First, the traversability
condition precludes the use of a matter system undergoing intense
nuclear processes in its interior. A very rough definition of a brown
dwarf is a ``star'' which forms via gravitational collapse but does
not ignite nuclear fusion in its core. Second, it is reasonable that
if such a system were to form there would be some cut-off point in the
stress-energy. Our system here possesses such a boundary (the stellar
surface) and therefore the analogy with stellar structure again leads
us to the term brown dwarf. It should be noted that $d_{+}$ does not
necessarily have to be equal to $d_{-}$ and therefore the brown dwarf
``star'' may have a different size in each universe.

\begin{figure}[ht]
\begin{center}
\includegraphics[bb=51 166 501 521, scale=0.7, keepaspectratio=true]{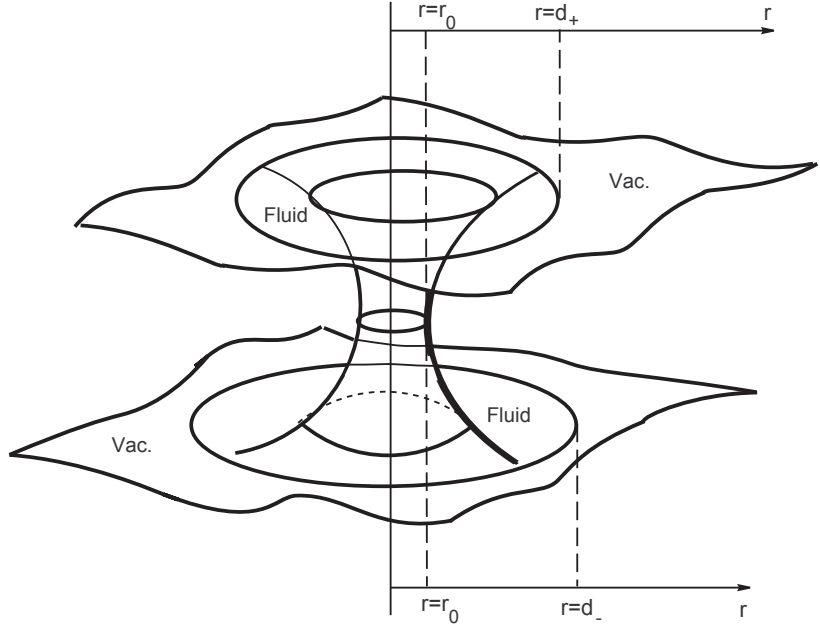}
\caption{{\small Embedding diagram for the wormhole geometry separating
two universes. Circles of constant $r$ represent two-spheres. There is
a throat at $r=r_{0}$ and the boundary of the brown dwarf is located
at $r=d_{\pm}$ where it is patched smoothly to Schwarzschild-(anti) de
Sitter geometry.}} \label{fig1}
\end{center}
\end{figure}

\qquad We take as our starting point the spherically symmetric line element of
the form
\begin{equation}
ds^{2}=-e^{\gamma(r)_{\pm}}\, dt^{2} + e^{\alpha(r)_{\pm}}\, dr^{2} + r^{2}\,
d\theta^{2} + r^{2}\sin^{2}(\theta)\, d\phi^{2} \label{eq:sphereline}
\end{equation}
with $\gamma(r)_{\pm}$ and $\alpha(r)_{\pm}$ being functions of the radial
coordinate only. The coordinates cover the range
\begin{equation}
-\infty < t <\infty,\;\; 0 < r_{0}< r < \infty,\;\; 0 < \theta <
\pi,\;\; 0\leq\phi < 2\pi .
\end{equation}
This form will prove especially useful when patching solutions to the
vacuum at $r=d_{\pm}$. The condition that there be no horizons implies
that $\gamma(r)_{\pm}$ be bounded \cite{ref:vishvesh}. Also, we demand
that the time-like
coordinate, $t$, be smooth across the throat which yields the
condition $\gamma(r_{0})_{-} = \gamma(r_{0})_{+}$.

\qquad The fundamental equations governing the geometry are the Einstein field
equations \footnote{Conventions here follow those of \cite{ref:MTW} with
$G=c=1$. The Riemann tensor is therefore given by $R^{\mu}_{\;\rho\nu\sigma} =
\Gamma^{\mu}_{\;\rho\sigma,\nu}+
\Gamma^{\mu}_{\;\alpha\nu}\Gamma^{\alpha}_{\;\rho\sigma} -\ldots$ with
$R_{\rho\sigma}=R^{\alpha}_{\;\rho\alpha\sigma}$.} with cosmological constant
\begin{equation}
R_{\mu\nu}-\frac{1}{2}R\,g_{\mu\nu} + \Lambda\,g_{\mu\nu}=8\pi T_{\mu\nu},
\label{eq:einsteq}
\end{equation}
along with the conservation law
\begin{equation}
T^{\mu\nu}_{\;\;\; ;\nu}=0. \label{eq:bianchi}
\end{equation}

\qquad The spherically symmetric Einstein field equations, in mixed form,
yield only three independent equations:
\begin{subequations}
\begin{align}
G^{0}_{\pm\; 0}=&\frac{e^{-\alpha(r)_{\pm}}}{r^{2}}\left( 1-r\alpha(r)_{\pm,1}
\right)-\frac{1}{r^{2}}= 8\pi T^{0}_{\pm\;0} - \Lambda \label{eq:einstzero} \\
G^{1}_{\pm\; 1}=&\frac{e^{-\alpha(r)_{\pm}}}{r^{2}}\left(
1+r\gamma(r)_{\pm,1}
\right)-\frac{1}{r^{2}}= 8\pi T^{1}_{\pm\;1} - \Lambda \label{eq:einstone} \\
G^{2}_{\pm\;2}=G^{3}_{\pm\;3}=& \frac{e^{-\alpha(r)_{\pm}}}{2} \left(
\gamma(r)_{\pm,1,1} +\frac{1}{2} \left(\gamma(r)_{\pm,1}\right)^{2}
+\frac{1}{r} \left(\gamma(r)_{\pm}-\alpha(r)_{\pm}\right)_{,1} \right.\nonumber \\
&\left.-\frac{1}{2}\alpha(r)_{\pm,1}\gamma(r)_{\pm,1} \right)=8\pi
T^{2}_{\pm\;2}-\Lambda = 8\pi T^{3}_{\pm\;3}-\Lambda, \label{eq:einsttwo}
\end{align}
\end{subequations}
whereas the conservation law yields one non-trivial
equation:
\begin{equation}
\frac{r}{2}T^{1}_{\pm 1,1}+
\left(1+\frac{r}{4}\gamma(r)_{,1}\right)T^{1}_{\pm 1}
-\frac{r}{4}\gamma(r)_{,1}T^{0}_{\pm 0} = T^{2}_{\pm 2}=T^{3}_{\pm 3}.
\label{eq:ttwotwo}
\end{equation}
Note that the cosmological constant, $\Lambda$, is assumed to have the
same value in both the $+$ and $-$ region. This is because the
cosmological constant is taken to represent the stress-energy of the
vacuum and we assume here that both regions, being connected, possess
similar vacua. It may be interesting to study situations where, for
example, one universe is of de Sitter type and the other of anti-de
Sitter type. In this case the de Sitter universe would be a ``baby universe''
to the anti-de Sitter one.

Equation (\ref{eq:einstzero}) may be utilized to give the following:
\begin{eqnarray}
e^{-\alpha(r)_{\pm}}&=&\frac{8\pi}{r}\int (r\pme)^{2} \left(T^{0}_{\pm
0} -\frac{\Lambda}{8\pi}\right) dr\pme +1 \nonumber \\ &=&
\frac{8\pi}{r} \int_{r_{0}}^{r}(r\pme)^{2} \left(T^{0}_{\pm 0} -
\frac{\Lambda}{8\pi}\right) dr\pme +\frac{c}{r} +1 =:
1-\frac{b(r)_{\pm}}{r}, \label{eq:alphaeq}
\end{eqnarray}
with $b(r)_{\pm}$ defined as the {\em shape function}. If
one is considering standard stellar models with plain topology, one
must set the constant $c=0$ to avoid a singularity at the origin.
Here, however, the origin ($r=0$) is not a part of the manifold (see
figure 1) and this constant must be set by the requirement that the
throat region be sufficiently smooth, as described next.

\qquad Figure 2 shows a cross-section or profile curve of a two
dimensional section, $t=\hbox{constant}$, $\theta=\pi/2$, of the
wormhole near the throat. The wormhole itself is constructed via the
creation of a surface of revolution when one rotates this figure about
the $x^{3}$-axis. The surface of revolution may be parameterized as
follows
\begin{equation}
x(r,\phi):=\left(x^{1}(r,\phi),\, x^{2}(r,\phi),\,x^{3}(r,\phi)
\right) = \left(r\cos(\phi),\, r\sin(\phi),\,P(r)\right),
\end{equation}
where $r=\sqrt{\left(x^{1}\right)^{2}+ \left(x^{2}\right)^{2}}$ and
$\phi$ is shown in figure 2. Therefore, the induced metric on the surface is given by
\begin{equation}
d\sigma^{2} = \left[1+P_{,1}(r)^{2}\right]\,dr^{2}+ r^{2}\,d\phi^{2},
\end{equation}
where commas denote ordinary derivatives. Note that the corresponding
four-dimensional spacetime metric, (\ref{eq:sphereline}), may now be written
as:
\begin{equation}
ds^{2}=-e^{\gamma(r)}\, dt^{2} + \left[1+P_{,1}(r)^{2}\right]\,dr^{2}+
r^{2}\,d\theta^{2}+r^{2}\sin^{2}(\theta)\,d\phi^{2}.
\label{eq:modsphereline}
\end{equation}
The shape function, $b(r)$, may now be specified in terms of the embedding
function, $P(r)$, which will be discussed shortly.

\begin{figure}[ht]
\begin{center}
\includegraphics[bb=106 196 491 586, scale=0.6,
keepaspectratio=true]{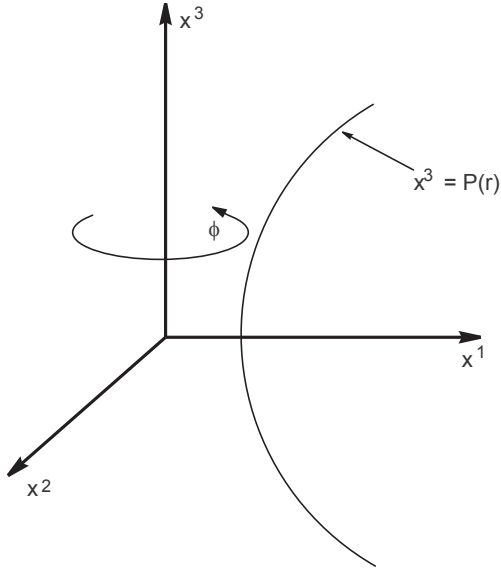} \caption{{\small A cross section
of the wormhole profile curve near the throat region. The wormhole is
generated by rotating the curve about the $x^{3}$-axis.}} \label{fig2}
\end{center}
\end{figure}

\qquad To set the constant, $c$, in (\ref{eq:alphaeq}) note that from the
profile curve in figure 2 the derivative $P_{,1}(r)$ tends to infinity
as one approaches $r_{0}$. This condition translates, via
(\ref{eq:modsphereline}), to $1-b(r)_{\pm}/r$ vanishing at $r=r_{0}$.
Therefore the constant $c$ must be equal to $-r_{0}$. This condition is
equivalent to that in \cite{ref:morthorne} and reflects the fact that
at the throat a finite change in coordinate distance, $\Delta r$,
corresponds
to an infinite change in proper radial distance, $\Delta l$. That is,
\begin{equation}
\lim_{r\rightarrow r_{0}} (\Delta l)= \lim_{r\rightarrow r_{0}}
\frac{\Delta r}{\sqrt{1-\frac{b(r)_{\pm}}{r}}}\rightarrow \infty .
\label{eq:grrblowup}
\end{equation}
A final condition for smooth patching is the continuity of the second
fundamental form \cite{ref:junction}. For the moment we assume that the
joining of the upper and lower universes may be performed. It will be
shown, in section 3, that the second fundamental form is indeed
continuous however; a few more properties of the system need to be
studied in order to demonstrate this.

\qquad We are now in a position to analyze the geometry. For now we limit
the analysis to the top half of the profile curve (the ``+'' region)
since the bottom half is easily obtained once we have the top half. From
the profile curve in figure 2 it may be seen that a function is needed
with the following properties:
\begin{itemize}
\item The derivative, $P_{,1}(r)$, must be infinite at the throat and
finite elsewhere.
\item The derivative, $P_{,1}(r)$, must be positive at
least near the throat.
\item The function $P(r)$ must possess a negative second derivative at
least
near the throat region.
\item Since there will be a cut-off at $r=d$, where the solution is joined
to vacuum (Schwarzschild-(anti) de Sitter), no
specification needs to be
made as $r$ tends to infinity.
\end{itemize}
These restrictions are purely geometric in nature. We show below
that reasonable physics will demand further restrictions. From these
properties a general function is assumed of the form:
\begin{subequations}
\begin{align}
P(r)&= A\left(r-r_{0}\right)^{\epsilon}\exp\left[-h(r)\right],
\label{eq:embedfunction} \\
P_{,1}(r)&=P(r)\left[\epsilon\left(r-r_{0}\right)^{-1}-h_{,1}(r)\right],
\label{eq:embedprime}  \\
P(r)_{,1,1}&=P(r)\left\{\left[\epsilon\left(r-r_{0}\right)^{-1}-
h_{,1}(r)\right]^{2}- h(r)_{,1,1} - \epsilon
\left(r-r_{0}\right)^{-2}\right\}. \label{eq:embeddprime}
\end{align}
\end{subequations}
Here $h(r)$ is any non-negative (at least twice differentiable)
function with non-negative first derivative. Moreover, $\epsilon$ is a
constant bounded as follows: $0 < \epsilon < \frac{1}{2}$. The reason
for this restriction will be made clear later. $A$ is a positive
constant which is required so that quantities possess correct units
(the unit of $A$ is $\hbox{length}^{1-\epsilon}$.) One may enforce
conditions on $h(r)$ by writing it in terms of a slack function as
\begin{eqnarray}
h_{,1}(r)&=&\hbox{e}^{q(r)}, \nonumber \\
h(r)&=&\int_{r_{0}}^{r} \hbox{e}^{q(r\pme)}\, dr\pme +z_{0}, \nonumber
\end{eqnarray}
with $z_{0}$ a constant and $q(r)$ an arbitrary differentiable function.

The shape function, $b(r)$, is given by (\ref{eq:alphaeq}) and (\ref{eq:modsphereline})
\begin{subequations}
\begin{align}
b(r)=&r\frac{P_{,1}(r)^{2}}{1+P_{,1}(r)^{2}}, \label{eq:shape} \\
b(r)_{,1}=& \frac{P_{,1}(r)}{1+P_{,1}(r)^{2}}\left[P_{,1}(r)+
\frac{2rP(r)_{,1,1}}{1+P_{,1}(r)^{2}}\right]. \label{eq:shapeprime}
\end{align}
\end{subequations}
The bottom universe may be obtained via the profile curve by demanding
that $A$ be some negative constant. The first derivative will then be
negative near the throat and the second derivative positive. The
function $h(r)$ may be different in the lower universe than in the
upper and this will not spoil smoothness at the throat as long as
$h_{+}(r_{0})=h_{-}(r_{0})$ and $h_{+,1}(r_{0})=h_{-,1}(r_{0})$.

\qquad To show that the system is indeed a wormhole with a smooth patching
at $r_{0}$, it is useful to show that the system may be described by a
single coordinate chart across the throat. Consider the profile curve in
the $x^{1}x^{3}$-plane of figure 2. The upper half of this curve may be
described by \begin{subequations} \begin{align}
x^{1}=&X_{+}^{1}(r_{+}):=r_{+} , \label{eq:x1+} \\
x^{3}=&X_{+}^{3}(r_{+}):=P_{+}(r_{+}), \label{eq:x3+} \end{align}
\end{subequations} with the $+$ subscripts indicating the upper universe
with $r_{+}>r_{0}$. We denote the inverse of $P_{+}(r_{+})$ from
(\ref{eq:x3+}) as $r_{+}=P_{+}^{\sharp}(x^{3})$ ($\sharp$ superscripts
will be used to indicate inverse mappings). Also, the lower half of the
curve is given by \begin{subequations} \begin{align}
x^{1}=&X_{-}^{1}(r_{-}):=r_{-} ,\label{eq:x1-} \\
x^{3}=&X_{-}^{3}(r_{-}):=P_{-}(r_{-}),\label{eq:x3-} \end{align}
\end{subequations} and again $r_{-}>r_{0}$ with
$r_{-}=P_{-}^{\sharp}(x^{3})$. At the moment we have two charts in two
universes and we may construct two metrics from these (the ``+'' and ``-''
metrics from above). The unifying chart, which includes both universes
near the throat as well as the boundary, $r_{0}$, is given by studying the
profile curve as a function of $x^{3}$. In other words, we rotate the
$x^{1}x^{3}$-plane about the $x^{2}$-axis of figure 2 by $\pi/2$ and
study the profile curve parameterized by the coordinate $x^{3}$. By
equation (\ref{eq:embedfunction}), $x^{3}$  belongs to an interval
containing $x^{3}=0$ which corresponds to the throat of the wormhole
($r_{0}$ corresponds to $P^{\sharp}(0)$). With
this parameterization, $P^{\sharp}(0)=r_{0}$ exists as well as
$\lim_{x^{3}\rightarrow 0_{+}}P^{\sharp}(x^{3}) = \lim_{x^{3}\rightarrow
0_{-}} P^{\sharp}(x^{3})$ and therefore $P^{\sharp}$ is continuous. The
functions $P^{\sharp}_{+}(x^{3})$ and $P^{\sharp}_{-}(x^{3})$ are simply
obtained by the restrictions of $P^{\sharp}(x^{3})$ in $x^{3} > 0$ and
$x^{3} < 0$ respectively. The resulting curve is parameterized as:
\begin{subequations}
\begin{align}
x^{1}=&X^{\sharp\;1}(x^{3})=P^{\sharp}(x^{3}) , \\
x^{3}=&X^{\sharp\;3}(x^{3}):=x^{3}.
\end{align}
\end{subequations}
With this ``universal'' chart the spacetime metric near the throat may be
written as
\begin{equation}
ds^{2}=-e^{\gamma^{\prime}(x^{3})}\, dt^{2}
+\left[1+P^{\sharp}_{,x^{3}}(x^{3})^{2}\right]\, (dx^{3})^{2} +
P^{\sharp}(x^{3})^{2}\, d\theta^{2} +P^{\sharp}(x^{3})^{2}\sin^{2}\theta\,
d\phi^{2}. \label{eq:newchart}
\end{equation}
This single metric is therefore sufficient to describe both parts of the
wormhole system across the throat.

\qquad In the coordinate chart used thorughout most of this paper,
(\ref{eq:modsphereline}), the manifold must be at least piece-wise $C^{3}$ 
smooth to have a well behaved Einstein tensor
with contracted
Bianchi identity. At the throat junction we will
later show that a minimum of $C^{1}$ is admitted although there is no
reason why a higher, perhaps more physical, class may not be admitted.
It may be argued that the chart given by (\ref{eq:newchart}) is better
suited for studies across the throat. In that case the embedding
function, $P^{\sharp}(x^{3})$, may be chosen of arbitrary
differentiability class (ideally minimum $C^{4}$ to give a $C^{3}$
manifold) and no piece-wise
considerations need be
addressed at the throat. The coordinate transformations from one chart
to the other, however, may turn out to be formidable. The above discussion
refers to the differentiability of the metric. Since the metric depends on
the derivative of the profile curve, a $C^{n}$ manifold should possess a
$C^{n+1}$ profile curve.

\section{Wormhole Structure}
\qquad In this section the field equations are utilized to construct
the wormhole structure. There are a number of ways to proceed and each
method has its particular advantages and disadvantages. One method is
a purely geometric method or g-method \cite{ref:syngebook}. In this
method one is free to specify the metric components, $\gamma(r)$ and
$b(r)$, to acquire the desired geometry. One then completely
determines the supporting stress-energy tensor from the geometry via
the field equations. This is the method of \cite{ref:morthorne}. One
may, for example, wish that $\gamma(r)$ be bounded to eliminate
horizons. As a simple example, such a function may be expressed in terms
of a sum of an even
and an odd bounded function as:
\begin{equation}
\gamma(r)=c_{1}\tanh(\varphi(r))+c_{2}\,\, \hbox{sech}(\psi(r)),
\end{equation}
which is bounded as
\begin{equation}
|\gamma(r)| \leq |c_{1}|+|c_{2}| .
\end{equation}
Here $\varphi(r)$ and $\psi(r)$ are arbitrary.
This method possesses the
advantage that any reasonable geometry allowed by general relativity
may be constructed. The disadvantage, however, is that one may be left
with a stress-energy tensor which is physically unreasonable.

\qquad A second method is the T-method \cite{ref:syngebook}. Here one
prescribes properties of a desired matter field and from this
constructs a stress-energy tensor. The field equations are then solved
for the metric which governs the geometry. This is generally a
preferred method as one may prescribe physically reasonable matter and
study the resulting gravitational field. The disadvantage, of course,
is that {\em there is very little control over the geometry}. If we have a
particular geometry in mind, there is no guarantee that an a priori
prescribed stress-energy tensor will generate the desired gravitational field.
Therefore, for the purposes of wormhole analysis, this method is not
useful.

\qquad A third method, and one which most of the analysis here is based
on, is a mixed method \cite{ref:das1}. This method may be implemented
in two ways:
\begin{itemize}
\item Prescribe some desirable physical properties to $T^{\:0}_{\pm 0}$
and $T^{\:1}_{\pm 1}$. For instance, for physically reasonable matter
one may wish the first to be negative and the latter non-negative. Otherwise,
$T^{\:1}_{\pm 1}$ may be freely set. $T^{\:0}_{\pm 0}$ may be obtained
via the shape function, $b(r)_{\pm}$, using (\ref{eq:einstzero}). The
physical properties demanded above must be utilized to constrain the
shape function. The transverse pressures $T^{\:2}_{\pm 2}=T^{\:3}_{\pm 3}$ are {\em defined}
via (\ref{eq:ttwotwo}).

\item Again obtain $T^{\:0}_{\pm 0}$ from the shape function as above and prescribe
an equation of state relating $T^{\:1}_{\pm 1}$ to $T^{\:0}_{\pm 0}$. Again the transverse
pressures are obtained from (\ref{eq:ttwotwo}).
\end{itemize}

All equations will then be satisfied. The advantage of this method is that
one may employ some reasonable physical assumptions on the matter
fields, yet not give up all control over the geometry. The major
disadvantage is that, since $g_{00}$ is not set, certain desirable
properties of the manifold, such as the absence of horizons and
curvature singularities, are not guaranteed. These problems will be
addressed below.

\qquad The matter field may, for example,  be that of an anisotropic fluid
whose stress-energy tensor is given by
\begin{equation}
T_{\mu\nu}=\left(\rho +
p_{\perp}\right)u_{\mu}u_{\nu}+p_{\perp}\,g_{\mu\nu}
+\left(p_{\shortparallel}-p_{\perp}\right)s_{\mu}s_{\nu}
\label{eq:setensor}
\end{equation}
with
\begin{subequations}
\begin{align}
u^{\mu}u_{\mu}=&-1, \\ s^{\mu}s_{\mu}=&1, \\ u^{\mu}s_{\mu}=&0.
\end{align}
\end{subequations}
Here, in the static case, the eigenvalues of $T^{\mu}_{\nu}$ are $T^{0}_{\; 0}=-\rho$,
$T^{1}_{\; 1}=p_{\shortparallel}$ and $T^{2}_{\;2}=T^{3}_{\;
3}=p_{\perp}$ which are a measure of the (negative) energy density, radial
pressure and transverse pressure, respectively. This matter model fits
our prescription for several reasons. First, the construction of
static, spherically symmetric wormholes requires a stress-energy
tensor which admits four real eigenvalues. Two of these eigenvalues
coincide by demanding spherical symmetry ($T^{2}_{\:2}$ and
$T^{3}_{\:3}$). Second, utilizing both the g and mixed methods will
most likely result in a system possessing three distinct eigenvalues.
Furthermore, it is known that these eigenvalues must, at least in some
vicinity, violate energy conditions \cite{ref:morthorne}. It has been
shown \cite{ref:daspaper} that the anisotropic fluid, when undergoing
gravitational collapse, may form many states of ``exotic'' (energy
condition violating) matter. Finally, a wormhole system as
described in this paper would most likely form via gravitational
collapse, as with normal stars. The anisotropic fluid therefore
represents a matter field which is a reasonable extension of certain
standard stellar models involving a perfect fluid \cite{ref:weinberg}.

\qquad The metric function , $\gamma_{\pm}(r)$, is given by
(\ref{eq:einstone}) and its integral:
\begin{subequations}
\begin{align}
\gamma(r)_{\pm ,1} =&
\left(8\pi T^{\:1}_{\pm 1}(r) -\Lambda +\frac{1}{r^2}\right)
\frac{r}{1-\frac{b_{\pm}(r)}{r}}-\frac{1}{r}, \label{eq:gammaprime} \\
\gamma(r)_{\pm} =&\int_{r_{0}}^{r} \left(8\pi T^{\:1}_{\pm 1}(r\pme)
-\Lambda
+\frac{1}{(r\pme)^{2}}\right)
\frac{r\pme\,dr\pme}{1-\frac{b_{\pm}(r\pme)}{r\pme}}-\ln(r) +
K_{\pm},
\label{eq:gamma}
\end{align}
\end{subequations}
with $K_{\pm}$ a constant. Continuity of $\gam$ at the
throat requires that $K_{+}=K_{-}$; $\alpha(r)_{\pm}$ is given by equation
(\ref{eq:alphaeq}):
\begin{subequations}
\begin{align}
\alp
=&-\ln\left[1-\frac{r_{0}}{r}+\frac{8\pi}{r}\int_{r_{0}}^{r}(r\pme)^{2}
\left(T^{0}_{\pm 0}-\frac{\Lambda}{8\pi}\right)\,dr\pme\right] \nonumber
\\
=&-\ln\left(1-\frac{b_{\pm}(r)}{r}\right), \label{eq:alphaequat} \\
\alpha(r)_{\pm ,1} =&\frac{1}{r\left(1-\frac{b_{\pm}(r)}{r}\right)}
\left(b(r)_{\pm,1}-\frac{b(r)_{\pm}}{r}\right).
\label{eq:alphaprime}
\end{align}
\end{subequations}
Another useful equation, which will be utilized later, is obtained by
subtracting (\ref{eq:einstzero}) from (\ref{eq:einstone})
\begin{equation}
8\pi e^{\alpha(r)_{\pm}} r \left(T^{1}_{\pm 1}-T^{0}_{\pm 0}\right)
=\left(\gamma(r)_{\pm} + \alpha(r)_{\pm}\right)_{,1}\;\;.
\label{eq:pplusrho}
\end{equation}

\qquad Having expressions for the metric functions and their
derivatives, we now turn to the matter field. This analysis will allow us
to determine all properties which the general embedding function
need to satisfy and will therefore provide a general class of metrics
describing such wormholes.

\qquad The Einstein equation (\ref{eq:einstzero}) may be written in
terms of the shape function yielding:
\begin{equation}
-\frac{1}{r^{2}}b(r)_{,1}=8\pi T^{\:0}_{\pm 0}- \Lambda.
\label{eq:bdensity}
\end{equation}
We assume here that, at least in the vicinity of the throat where the
cosmological constant is negligible compared to the matter energy
density, the net energy density is positive. From (\ref{eq:bdensity})
it is seen that this requires that $b(r)_{,1} > 0$ near the throat.
Recall from the properties of the profile curve (figure 2), $P(r)$
must possess positive first derivative and negative second derivative
near the throat. From (\ref{eq:shapeprime}) it may be seen that if the
second term in the expression dominates, $b(r)_{,1}$ will be negative
which is not desirable. Therefore (\ref{eq:shapeprime}) must be
analyzed with some care in the vicinity of the throat.

\qquad To study the properties of (\ref{eq:bdensity}) near the throat,
we define a parameter $\delta:=(r-r_{0})$. Analysis of $b(r)_{,1}$ then
shows that
\begin{equation}
b(r)_{,1}\approx 1-\mathcal{O}(\delta^{1-2\epsilon}) + \ldots
\label{eq:approxbprime}
\end{equation}
from which it may be seen that $\epsilon$ must be restricted to the
interval $0 < \epsilon < \frac{1}{2}$, as mentioned earlier. Note that
$\epsilon =\frac{1}{2}$ is allowed, although a negative energy density
as observed by static observers is the price one pays for this case.
Even for the values of $\epsilon$ stated above $b(r)_{,1}$ eventually
becomes negative at some radius, $r=r_{\pm}^{*}$ say, as is evident
from (\ref{eq:shapeprime}). One could, if one insists on a positive
value of $b(r)_{,1}$ everywhere, place the stellar boundary at $d_{\pm}<
r_{\pm}^{*}$. Note that since the model admits an arbitrary cosmological
constant, we could allow $b(d_{\pm})_{,1}<0$ and still claim a positive
$\rho$ for the matter field if $\Lambda < 0$.

\qquad The equation (\ref{eq:approxbprime}) with the above restriction on
$\epsilon$ yields the best possible scenario as far as the energy density
is concerned. That is, a value of $b(r)_{\pm ,1}$ greater that unity
cannot be achieved at the throat. This situation is quite favourable
with respect to the energy conditions (to be discussed section 4) as the null, weak and
dominant energy conditions each require
positivity of energy density.

\subsection{Horizons and Singularities}
\qquad As mentioned earlier, the method utilized here, although giving
some control over the matter fields, does not guarantee the absence of
event horizons nor singularities. In this section we address both of
these issues. The notation $\tau(r):=8\pi T^{1}_{\;1}-\Lambda$ is used
throughout this section.

\qquad In the coordinate system of (\ref{eq:sphereline}) an event
horizon will exist wherever $g_{00}$ vanishes. In other words, the
function $\gamma_{\pm}(r)$ must be everywhere finite. Using
(\ref{eq:gammaprime}) and (\ref{eq:gamma}) it may be seen that, to
ensure that $\gam$ remain finite near the throat, $(\tau(r)+1/r^{2})$
must vanish at least as fast as $(r-b(r)_{\pm})$ in the limit
$r\rightarrow r_{0}$.

\qquad From (\ref{eq:shape}) it may be shown that near the throat
\begin{equation}
\left(b(r)-r\right)\propto \left(r-r_{0}\right)^{2(1-\epsilon)};
\label{eq:bnearthroat}
\end{equation}
therefore the numerator of the first term in (\ref{eq:gammaprime}) must possess behaviour
near the throat of the
form $(r-r_{0})^{\eta}$ with $\eta \geq 2(1-\epsilon)$ and
\begin{equation}
\tau(r_{0})=-1/r_{0}^{2}, \label{eq:tauthroat}
\end{equation}
Although we may possess everywhere non-negative energy density,
(\ref{eq:tauthroat}) shows that
a
negative pressure or {\em tension} is required at the throat to
support the wormhole. The fluid pressure may still be positive if the
cosmological constant is positive and dominates the fluid pressure at
the throat.

\qquad The equation (\ref{eq:gamma}) may most easily be made
well behaved for any $\epsilon$ allowed by the model by demanding that
 the first-order term in the expansion of $\tau(r)+1/r^{2}$ about
$r_{0}$ vanish. Also, the appropriate limit $\tau(r_{0})=-1/r_{0}^{2}$ must
be enforced so that the zeroth-order term vanish as well. With little loss
of generality we may write $\tau(r)$ as
\begin{eqnarray}
\tau(r)&=&-\frac{1}{r^{2}}
\cos\left[\frac{(r-r_{0})}{(d+\alpha-r_{0})}\frac{m\pi}{2}\right]
\cosh\left[(r-r_{0})^{s}\xi(r)\right] \nonumber \\
&+&\kappa (r-r_{0})^{q}(r-d)^{w}\exp\left(\zeta(r)\right)
\label{eq:tauansatz}
\end{eqnarray}
with $m$ an odd integer, $q\geq 2, w>0$ positive integers and $\xi(r)$ and $\zeta(r)$
arbitrary
functions save for the restriction that they possesses well behaved
derivatives in the regime $r_{0}\leq r \leq d$; $\alpha$ is a small
constant set by the requirement that at $r=d_{\pm}$ the above expression
is equal to $-\Lambda$. The second term, proportional to the constant
$\kappa$, is not strictly required, though it is useful in making
$\tau(r)$
positive throughout most of the stellar region (see section 4). An analysis
of the matter field in the context of the energy conditions may also be found
in section 4.

\qquad Having eliminated the possibility of event horizons, it is now necessary to
analyze the manifold for singularities. Singular manifold structure is most easily studied by
constructing the Riemann tensor in an orthonormal frame. We use notation such that
indices surrounded by parentheses denote quantities in the orthonormal
frame. In this frame the Riemann tensor possesses the following components
as well as those related by symmetry \cite{ref:morthorne}:
\begin{subequations}
\begin{align}
R_{(trtr)\pm}=& \frac{1}{2}\alphab\left[ \gamma(r)_{\pm,1,1}
+\frac{1}{2}(\dgam)^{2} \right. \nonumber \\
&- \left.
\frac{1}{2}\dgam
\frac{b(r)_{\pm,1}-\frac{b_{\pm}(r)}{r}}{r-b_{\pm}(r)}\right]\;\; ,
\label{eq:riem1} \\
R_{(t\theta t\theta)\pm} =& \frac{1}{2r}\dgam\alphab = R_{(t\phi t\phi)\pm}\;\; ,
\label{eq:riem2} \\
R_{(r\theta r\theta)\pm} =& \frac{1}{2 r^{2}}\left[b(r)_{\pm,1}-
\frac{b_{\pm}(r)}{r}\right] = R_{(r\phi r\phi)\pm}\;\; , \label{eq:riem3} \\
R_{(\theta\phi\theta\phi)\pm} =& \frac{b_{\pm}(r)}{r^{3}}\;\; .
\label{eq:riem4}
\end{align}
\end{subequations}
Recall that, utilizing our ansatz (\ref{eq:embedfunction}) with
(\ref{eq:shape}) and (\ref{eq:shapeprime}), $b(r_{0})=r_{0}$ and
$b(r_{0})_{,1}=1$. Therefore all components of the Riemann tensor, with
the possible exception of (\ref{eq:riem1}), are finite at the throat.
The only other problematic spot is $r=0$, which is not a part of the
manifold and therefore causes no concern for our analysis. The previous
restrictions on $\tau(r)_{\pm}$ ensure that $\gamma(r)_{\pm}$ and its
derivatives are finite away from the throat. No quantity grows without
bound as we move away from the throat.

\qquad To study (\ref{eq:riem1}) more carefully, it is
useful to write it
in terms of $T^{\:1}_{\pm 1}$ using (\ref{eq:gammaprime}) as (dropping the
$\pm$ subscript for now)
\begin{eqnarray}
R_{(trtr)}&=&\frac{1}{4 r^{3}\left(r-b(r)\right)}\left[ \tau(r)^{2}r^{6}
+2r^{4}\tau(r)_{,1}\left(r-b(r)\right) \right. \nonumber \\
&+&r^{3}\tau(r)\left(2r-b(r)+r b(r)_{,1}\right) -4rb(r)+2r^{2}b(r)_{,1}
\nonumber \\
&+&\left. 4b(r)^{2}-b(r)b(r)_{,1}r\right]. \label{eq:taurtrtr}
\end{eqnarray}
Notice that for non-singular behaviour, the term in square brackets
must vanish at least as fast as the denominator as $r\rightarrow
r_{0}$. Recall that the denominator's behaviour is as in
(\ref{eq:bnearthroat}) and therefore the numerator must possess
similar or stronger vanishing properties here. An exhaustive
calculation using (\ref{eq:tauansatz}) reveals that the numerator does
indeed vanish at least as fast as the denominator near the throat and
therefore all curvature tensor components are finite, eliminating any
singularities. It is interesting to note that, by demanding that
$\gamma(r)_{\pm ,1}$ be finite everywhere (the horizon condition), we
also impose the final condition which ensures finiteness on all
orthonormal Riemann tensor components. In other words, naked
singularities are {\em not} supported.

\qquad We now return to the consideration regarding the continuity of
the second fundamental form at the throat. The second form is most
easily calculated in the coordinate frame as the covariant derivative
\begin{equation}
K_{\mu\nu}=\frac{1}{2}\left(\hat{n}_{\mu;\nu}+\hat{n}_{\nu;\mu}\right),
\label{eq:extcurve}
\end{equation}
where $\hat{n}_{\mu}$ represents an outward pointing radial unit vector
which, using (\ref{eq:modsphereline}), is given by:
\begin{equation}
\hat{n}_{\mu}=\sqrt{1+P_{,1}(r)^{2}}\;\;\delta^{(r)}_{\;\mu}
\end{equation}
(the $\pm$ subscripts have been dropped here). From (\ref{eq:extcurve})
the following components are calculated for the second fundamental
form:
\begin{subequations}
\begin{align}
K_{tt}=&-\frac{1}{2\sqrt{1+P_{,1}(r)^{2}}}\hbox{e}^{\gamma(r)}\gamma_{,1}(r)\;
,
\label{eq:exttt} \\
K_{\theta\theta}=&\frac{r}{\sqrt{1+P_{,1}(r)^{2}}}\; ,
\label{eq:extthetatheta} \\
K_{\phi\phi}=&\frac{r \sin^{2}\theta}{\sqrt{1+P_{,1}(r)^{2}}}\; .
\label{eq:extphiphi}
\end{align}
\end{subequations}
(\ref{eq:extthetatheta}) and (\ref{eq:extphiphi}) are certainly
continuous at the throat since, from (\ref{eq:embedprime}) they both
vanish as one approaches $r=r_{0}$ from either the ``+'' region or the
``-'' region. $K_{tt}$ will be studied with some care. Recall that the
condition ensuring the absence of horizons is given by demanding that
(\ref{eq:gammaprime}) be everywhere finite. This condition therefore
requires that $\tau(r_{0}):=8\pi T^{1}_{\;1}(r)-\Lambda=-1/r_{0}$ and
this result is, of course, independent of the region in which one is
approaching $r_{0}$. This condition, along with the fact that
$b_{+}(r_{0})=b_{-}(r_{0})=r_{0}$ dictates, via (\ref{eq:gammaprime}),
that $\gamma_{,1}(r)$ be continuous at the throat. Also, continuity of
the metric requires that
$\hbox{e}^{\gamma(r_{0})_{+}}=\hbox{e}^{\gamma(r_{0})_{-}}$ and
therefore $K_{tt}$ must also be continuous at the throat. All required
junction conditions at the throat have therefore been met (namely:
continuity of the metric, $T^{1}_{\;1}(r)$, and the second fundamental
form as well as condition (\ref{eq:grrblowup}) along with visual proof
from figure 4b to be discussed in section 4).

\qquad If one is considering a traversable type wormhole then a
further restriction on Riemann components is generally desired.
Namely, the magnitude of these components must be small in the sense
that tidal acceleration, given  by:
\begin{equation}
\Delta a^{\mu}=R^{\mu}_{\; \alpha\beta\gamma}V^{\alpha}V^{\beta}\Delta\xi^{\gamma}
\label{eq:tidalaccel}
\end{equation}
be sufficiently small to allow travel through the throat. Here
$\Delta\xi^{\gamma}$ is a separation vector joining two parts of the
traveling body. The condition that (\ref{eq:tidalaccel}) be small
places constraints on the magnitude of the orthonormal Riemann tensor
components as well as on the velocity at which the wormhole may safely
be traversed \cite{ref:MTW}, \cite{ref:visbook}. At the throat, it is
possible to make all components of (\ref{eq:riem1})-(\ref{eq:riem4})
small or vanish due to the fact that $b(r_{0})=r_{0}$ and
$b(r_{0})_{,1}=1$, with possible exception of (\ref{eq:riem4}) which
is equal to $1/r_{0}^{3}$. For purely radial motion
(\ref{eq:tidalaccel}) dictates that this component plays no part in
the tidal acceleration and therefore presents no constraint on the
wormhole size \cite{ref:MTW}, \cite{ref:visbook}. Away from the throat
there is the potential to make the relevant components at least
reasonably small by the choice of matter field via
(\ref{eq:gammaprime}) and (\ref{eq:taurtrtr}). Note that the
``smallness'' of the Riemann components is not a requirement of the
wormhole system. It is simply a measure of the ease at which one may be
traversed.

\subsection{The Stellar Boundary}
\qquad Having identified conditions in the interior we now turn
attention to the boundary of the star at $r=d_{\pm}$. Here the
solution must smoothly tend to the vacuum Schwarzschild-(anti) de
Sitter metric:
\begin{eqnarray}
ds^{2}=&-&\left(1-\frac{2M_{0\pm}}{r}-\frac{1}{3}\Lambda\,r^{2}\right)d\tilde{t}^{2}
+\frac{dr^{2}}{\left(1-\frac{2M_{0\pm}}{r}-\frac{1}{3}\Lambda\,r^{2}\right)}
\nonumber \\ &+&r^{2}\,d\theta^{2} + r^{2}\sin^{2}(\theta)\,d\phi^{2}
\label{eq:schwds}
\end{eqnarray}
where the tilde on the time coordinate will be later made clear. Note
that the ``mass'' of the star in one universe need not necessarily be
the same as in the other. The junction condition employed at this
surface will be that of Synge \cite{ref:syngebook}
\begin{equation}
\left. T^{\mu}_{\pm\nu}n^{\nu}\right|_{r=d_{\pm}}=0 \label{eq:syngcond}
\end{equation}
with $n^{\nu}$ an outward pointing radial normal vector. This condition
essentially states that there may be no flux of stress-energy off the
surface of the star which, of course, is a reasonable requirement when
patching to a vacuum solution as will be done next. For the class of
metrics considered here, continuity of the metric at the boundary also
ensures continuity of the extrinsic curvature.

\qquad It is now necessary to check conditions at the boundary
$r=d_{\pm}$ where the interior solution must match the vacuum solution
given by (\ref{eq:schwds}). Here, from (\ref{eq:alphaeq}) we get
\begin{equation}
e^{\alpha(d_{\pm})}=\left[1-\frac{2M_{\pm}}{d_{\pm}}+\frac{\Lambda}{3}
\frac{r_{0}^{3}}{d_{\pm}}
-\frac{r_{0}}{d_{\pm}}-\frac{\Lambda}{3}d_{\pm}^{2}\right]^{-1}, \label{eq:alphaatd}
\end{equation}
where $M_{\pm}$ corresponds to the fluid mass,
\begin{equation}
M_{\pm}=-4\pi\int_{r_{0}}^{d_{\pm}}(r\pme)^{2}T^{0}_{\pm\;0}\,dr\pme .
\label{eq:fluidmass}
\end{equation}
The function (\ref{eq:alphaatd}) matches smoothly to the Schwarzschild-(anti) de Sitter
metric (\ref{eq:schwds}) with an ``effective mass''
\begin{equation}
M_{0\pm}=M_{\pm}-\frac{\Lambda}{6}r_{0}^{3}+\frac{r_{0}}{2}.
\label{eq:effectivemass}
\end{equation}
The equation for $\gamma_{\pm}$ is more complicated:
\begin{eqnarray}
e^{\gamma(d_{\pm})}&=&\left(1-\frac{2M_{0\pm}}{d_{\pm}}-\frac{\Lambda}{3}d^{2}_{\pm}\right)
\nonumber \\
&\times& \exp\left\{8\pi\int_{r_{0}}^{d_{\pm}}
\frac{(r\pme)^{2}(p_{\pm\smp}+\rho_{\pm})\;dr\pme}
{\left[r\pme-8\pi\int_{r_{0}}^{r\pme}(r^{*})^{2}(\rho_{\pm}+\frac{\Lambda}{8\pi})\,dr^{*}-
r_{0}\right]+ K}\right\} \nonumber \\ &=:&
\left(1-\frac{2M_{0\pm}}{d_{\pm}}-\frac{\Lambda}{3}d^{2}_{\pm}\right)e^{2\chi_{\pm}}.
\label{eq:gammaboundary}
\end{eqnarray}
Here (\ref{eq:pplusrho}) has been used. We now perform the following
coordinate change
\begin{equation}
\tilde{t}=e^{\chi}t \label{eq:ttrans}
\end{equation}
so that
\begin{equation}
\left(d\tilde{t}\right)^{2}= e^{2\chi}dt^{2}
\end{equation}
and thus the interior solution has been matched to (\ref{eq:schwds}).

\section{Examples}
\qquad Having developed the formalism for constructing such wormholes it
is instructive to demonstrate how the scheme works via specific examples. Again the $\pm$
subscript is neglected unless needed for clarity.

\qquad A simple, yet instructive example is
one with the following parameters:
\begin{subequations}
\begin{align}
h(r)\propto& \;\; r^{2}\:, \nonumber \\
\xi(r)\propto& \;\; r^{3}\:, \nonumber \\
m=&\;\; 3, \nonumber \\
s=&\;\; 1, \nonumber \\
\Lambda\sim &\;\; \hbox{small and positive}, \nonumber \\
\kappa\sim &\;\; \hbox{large and positive}, \nonumber \\
\zeta(r) &\;\; \hbox{a constant}. \nonumber
\end{align}
\end{subequations}
The rotated profile curve for this system is shown in figure 3a. Figure 3b
displays the shape function and 3c its derivative. Note that
the total mass (as measured by $b(r)$) and the energy density
($b(r)_{,1}$) are positive in the vicinity of the throat. Since the
Einstein equations permit discontinuities in the energy density, there is
no reason to place the stellar boundary near the point where $b(r)_{,1}$
vanishes although for this example we do so. The radial pressure,
$p_{\smp}=\left(\tau(r)+\Lambda\right)/8\pi$, is plotted in figure 3d.
Note
that with the addition of the large $\kappa$ term this term is positive
throughout most of the region. Finally, figure 3e shows a plot of the
transverse pressure. As with the radial pressure, this quantity is
positive throughout most of the stellar bulk.

\begin{figure}[ht]
\begin{center}
\includegraphics[bb=91 141 481 724, scale=0.45, keepaspectratio=true]{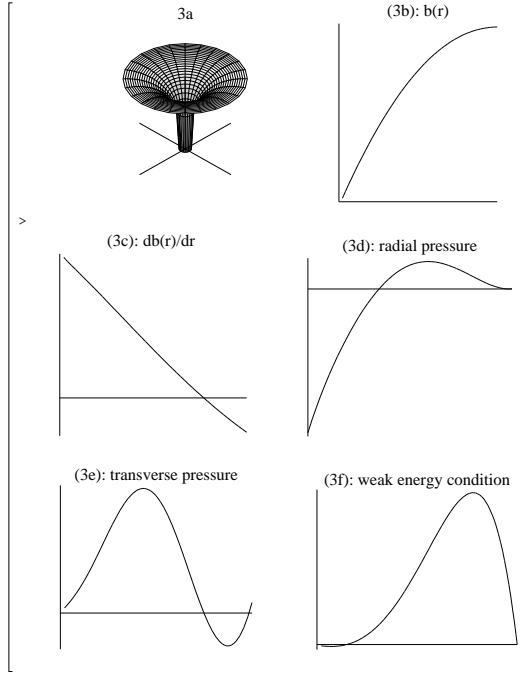}
\caption{{\small A sample wormhole system. a) shows a rotated profile curve of the ``+''
universe. b) is a plot of the effective mass of the star, c) a plot of the effective energy
density, $b(r)_{,1}$, d) the parallel pressure e) the transverse pressure and f) the weak
energy condition. Note that
$\rho$ is non-negative. Moreover, throughout most of the stellar bulk $p_{\smp}$ and
$p_{\perp}$ are positive.}}
\label{fig3}
\end{center}
\end{figure}

\subsection{Energy Conditions}
\qquad Here we discuss energy conditions both in general and with respect
to the above example. From an expansion of (\ref{eq:bdensity}) near the
throat we find that
\begin{eqnarray}
8\pi\rho&\approx& \frac{1}{r_{0}^{2}}-\Lambda +\frac{2}{A^{2}\epsilon
r_{0}}\hbox{e}^{2h(r_{0})}\left(1-\frac{1}{\epsilon}\right)
\left(r-r_{0}\right)^{1-2\epsilon}
-\frac{2}{r_{0}^{3}}\left(r-r_{0}\right) \nonumber \\
&-&\frac{1}{A^{2}\epsilon^{2}r_{0}^{2}}\hbox{e}^{2h(r_{0})}
\left(r-r_{0}\right)^{2(1-\epsilon)}.
\label{eq:rhonearthroat}
\end{eqnarray}
A similar analysis on the radial pressure yields via (\ref{eq:tauansatz}):
\begin{equation}
8\pi p_{\smp}\approx -\frac{1}{r_{0}^{2}}+\Lambda
+\frac{2}{r_{0}^{3}}(r-r_{0}) +\mathcal{O}(r-r_{0})^2.
\label{eq:Pnearthroat}
\end{equation}
Finally, the transverse pressure (\ref{eq:ttwotwo})
\begin{equation}
8\pi p_{\perp}(r_{0})=8\pi\frac{r_{0}}{2}p_{\smp ,1}|_{r=r_{0}}-\frac{1}{r_{0}^2}+\Lambda =
\Lambda .
\label{eq:perpnearthroat}
\end{equation}
Exactly at the throat, energy conditions are respected as $\rho$ is
positive unless the cosmological constant is large (a possibility ruled
out by experiment) and $\rho(r_{0})+p_{\smp}(r_{0}) = 0$ as well as
$\rho(r_{0})+p_{\perp}(r_{0}) > 0$. As we move away from the throat notice that, in the
expression $\rho + p_{\smp}$, the third and fifth terms of (\ref{eq:rhonearthroat})
survive. Both of these terms yield a negative contribution and therefore energy conditions are
violated slightly away from the throat. It is possible to cut-off the
solution at
some radius near $r_{0}$ and patch to an
intermediate
layer of material which respects energy conditions. This
intermediatelayer may then be
patched to the Schwarzschild-(anti) de Sitter vacuum and thus there will
be little or {\em no}
energy
condition violation. Notice from figure 3 that {\em averaged} energy conditions (energy
conditions integrated over the stellar bulk) are certainly satisfied as most of the stellar
bulk is ``well behaved'' in the context of energy conditions. Also interesting is that, at the
throat, the
matter is still exotic in the sense that the magnitude of the pressure is of the same order of
magnitude as the energy density. A positive cosmological constant, if present, would minimize
the amount of material needed at the throat although the cosmological constant would most
likely be quite
small when compared to $1/r_{0}^2$ and therefore would not necessarily eliminate this exotic
behaviour.

\qquad The weak energy condition may also most easily be studied by
forming the quantity \cite{ref:morthorne}, \cite{ref:delgaty}
\begin{equation} \varepsilon:=\frac{(p_{\smp}+\rho)}{|\rho|}.
\label{eq:weakcond} \end{equation} This quantity should be positive for
any matter respecting the weak energy condition. We plot this quantity for
the example above in figure 3f. This figure shows that throughout the
stellar bulk, with the exception of a small region near (but not at) the
throat, the weak energy condition is satisfied. The violation, however, is
very small. This agrees with the perturbative analysis above, as well as
with \cite{ref:kuhfittig}, where it si shown that the energy violations
can be confined to an arbitrarily small region around the throat provided
that $b_{1}(r)$ is close to unity. As pointed
out in much of the literature on energy conditions, this violation does
not conflict with experimental results as the weak energy violating
Casimir effect has been verified experimentally \cite{ref:casimir}.

\qquad Another interesting example, which we will briefly consider, is
that where the wormhole connects not just two universes but any number of
universes. An in-depth analysis of this situation will not be made and
this example simply serves to illustrate that a minor extension to the
above techniques is all that is required for such a system.

\qquad The profile curve of the ``top half'' of a single closed universe
is shown in figure 4a. The bottom half of the universe may be generated by
reflection via $P(r) \rightarrow -P(r)$. We use as an example of such a
profile curve the following function:
\begin{equation}
P(r)=\frac{1}{k}\cos^{-1}\left[\frac{(r-B)}{R}\right],
\label{eq:multiprofile}
\end{equation}
where $k$, $R$ and $B$ are constants (with $B>R$) related
to the size
of the universe and to the size of the throat via
$d=B+R$ and $r_{0}=B-R << B+R$. This profile, along with its reflection
about the
$r$-axis
serves to generate the closed universe. Note that this universe may be
joined at its throats to other (potentially infinite number of) closed
universes or to an open universe at each end of the closed universe
chain. In this latter case the closed system represents a set of ``baby
universes'' to the open ones. It should also be possible to terminate
the chain not with open universes but instead a closed universe with
only one throat. Note that continuity of the metric must be respected
at each coordinate chart junction in order to have a smooth patching.
A representative system of three closed universes with similar radii is
shown in figure 4b.

\begin{figure}[ht]
\begin{center}
\includegraphics[bb=71 342 455 733, scale=0.6,
keepaspectratio=true]{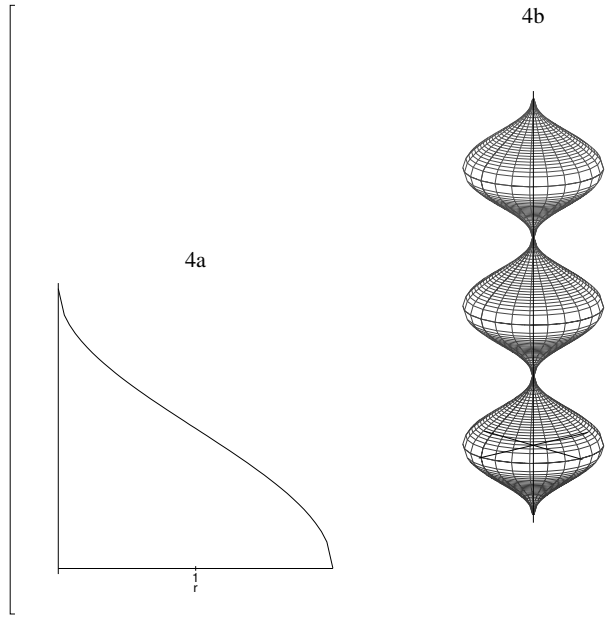} \caption{{\small a) The profile curve for the upper
section of a single universe. b) A chain of three closed universes connected by wormholes.}}
\label{fig4}
\end{center}
\end{figure}

\qquad Again we utilize the mixed method of solving the Einstein field
equations in order to minimize energy condition violation. Singularities
and event horizons are still forbidden and, since here we do not patch to
a well behaved vacuum solution, a slight modification of the above
techniques will be required to ensure these conditions both at the throat
and at $r=d:=(B+R)$.

\qquad The Einstein equations give, for the energy density of the
fluid,
\begin{eqnarray}
8\pi\rho + \Lambda &=&\frac{1}{r^{2}}b(r)_{,1} \nonumber \\
&=& \frac{k^{2}\left(R^2+r^{2}-B^{2}\right)+1}{
r^{2}\left[k^{2}\left((B-r)^{2}-R^{2}\right)-1\right]^{2}}
\label{eq:multirho}
\end{eqnarray}
where the $\pm$ subscript has been dropped as it is implied that
quantities in the top half of the universe should be equal to those in
the bottom half. Although this condition is not a strict requirement
of the model, it is used here to simplify the system. It should be noted
that, as mentioned above, the metric must be continuous at both at the
throat and at $r=d$ in
order to have a smooth patch. Also, the junction conditions require $T^{1}_{\;1}(d)$ in
the lower half to equal $T^{1}_{\;1}(d)$ in the upper half.

\qquad Equation (\ref{eq:multirho}) may be positive or negative. If one
wishes to demand a positive total energy density everywhere, then the
condition $r_{0}\leq\frac{1}{2k^{2}R}$ must hold. However, the presence of
a negative cosmological constant allows (\ref{eq:multirho}) to be
negative and still have everywhere non-negative energy density for the
fluid.

\qquad We now analyze the manifold for event horizons and
singularities. In this example, the Einstein equations give:
\begin{equation}
\gamma(r)_{,1}=\left(\tau(r)+\frac{1}{r^{2}}\right)r
\left[1+\frac{1}{k^{2}\left[R^{2}-(r-B)^{2}\right]}\right]-\frac{1}{r},
\label{eq:multidgamma}
\end{equation}
where the previous notation is still used $\left(\tau(r):=8\pi T^{1}_{\;1}
-\Lambda\right)$. Note that now there are two potentially troublesome
spots,
namely $r=r_{0}$ and $r=d$. As before, to guarantee that the integral on
(\ref{eq:multidgamma}) is well behaved, it is required that $\tau+1/r^{2}$
vanish at least as fast as the denominator of the second term in square
brackets at these spots. Again, as is common with static wormhole
solutions, a net tension is required at the throat to maintain the
wormhole. A tension is also required at $r=d$. Although many functions
will possess this behaviour, for the specific example considered here we
assume
\begin{equation}
\tau(r)=-\frac{1}{r_{0}^{2}}
\cos\left(\frac{r-r_{0}}{d-r_{0}}\frac{m\pi}{2}\right)-
\frac{1}{d^{2}}\cos\left(\frac{d-r}{d-r_{0}}\frac{m\pi}{2}\right),
\label{eq:multitau}
\end{equation}
where $m$ is an odd integer.

\qquad The Riemann curvature tensor in the orthonormal frame at
$r=r_{0}$ gives, from (\ref{eq:riem1}):
\begin{eqnarray}
R_{(trtr)}(r_{0})&=& \frac{1}{4 r_{0}^{3}\left(r_{0}-b(r_{0})\right)}
\left\{\tau^{2}(r_{0})r_{0}^{6}+\tau(r_{0})r_{0}^{4}\left(1+
b(r_{0})_{,1}\right)\right. \nonumber \\
&+&\left. r_{0}^{2}b(r_{0})_{,1}\right\}. \label{eq:multirtrtro}
\end{eqnarray}
By noting that $b(r_{0})=B-R$ and $b(r_{0})_{,1}=2k^{2}R(R-B)+1$ it is
easy to check that (\ref{eq:multitau}) sets the quadratic in the numerator
of the above expression to zero. Similarly, at $r=d$
\begin{eqnarray}
R_{(trtr)}(d)&=& \frac{1}{4 d^{3}\left(d-b(d)\right)}
\left\{\tau^{2}(d)r_{0}^{6}+\tau(d)d^{4}\left(1+
b(d)_{,1}\right)\right. \nonumber \\
&+&\left. d^{2}b(d)_{,1}\right\}. \label{eq:multirtrtd}
\end{eqnarray}
Here $b(d)=B+R$ and $b(d)_{,1}=2k^{2}R(R+B)+1$ and again
(\ref{eq:multitau}) gives a finite result. Finally, it should be noted
that, since $\rho$, $\tau(r)$ and $\gamma(r)_{,1}$ are everywhere finite, the transverse
pressures (defined via equation
(\ref{eq:bianchi})) are also well behaved.

\section{Conclusion} \qquad In this paper a general model of a static wormhole system was
developed. The construction was designed to encompass a large class of
static wormhole solutions and limitations were imposed on it so that
physical quantities are as reasonable as possible. The system was
shown to obey both energy conditions at the throat as well as averaged
energy conditions throughout the matter bulk. Also, the matter
system smoothly joins a vacuum possessing an arbitrary cosmological
constant. Many specific examples may be constructed from the work here
and a couple were considered, including a system of multiple connected
universes. It would be interesting to derive a general class of
time-dependent wormhole geometries, although the equations governing such
a
system are formidable. In this case one would also need to specify, at
the very least, some qualitative features of the future evolution of
the system. Examples are cases where the
wormhole effectively ``closes up'' or simply translates through
spacetime. As well there is the issue of causality (\cite{ref:mty},
\cite{ref:guts}, \cite{ref:kras}, \cite{ref:konst} and others). There
is also much interesting work to be done regarding wormholes of other
symmetries (for example see \cite{ref:gonzalez}, \cite{ref:aros}). It
should be straight-forward to apply the analysis here to generate a
general class of cylindrical wormholes, for example.

\qquad It is hoped that this work may provide a general wormhole model
from which one may base future work in this fascinating field. The aim
is also to shed light on methods for satisfying all field equations and
identities in a way relevant to studies of stellar systems.

\newpage
\bibliographystyle{unsrt}

\end{document}